\documentclass{amsart}
\usepackage{amssymb,latexsym}
\usepackage{graphicx}

\begin{document}

\title{Special polynomials and elliptic integrals}
\author{D. Babusci$^\dag$, G. Dattoli$^\ddag$}
\address{$^\dag$ INFN - Laboratori Nazionali di Frascati, via E. Fermi 40, I-00044 Frascati.}
\address{$^\ddag$ ENEA - Dipartimento Tecnologie Fisiche e Nuovi Materiali, Centro Ricerche Frascati\\
                 C. P. 65, I-00044 Frascati.}
\begin{abstract}
We show that the use of generalized multivariable forms of Hermite polynomials provide 
an useful tool for the evaluation of families of elliptic type integrals often 
encountered in electrostatic and electrodynamics
\end{abstract}

\maketitle

In this letter we introduce a technique, based on the use of generalized families of polynomials and integral 
transforms, to evaluate integrals, finite and infinite, including also hyper-elliptic forms. We introduce the method by 
considering, as an example, the following integral:
\begin{equation}
F(a,0|\nu,2) \,=\, \int_{-\infty}^{\infty}\, {\rm d}x\, \frac{1}{(1 + a x^2)^{\nu}}  \qquad\qquad (\Re\,\nu > \frac{1}{2},\,a > 0),
\end{equation}
whose explicit expression is easily obtained in terms of the Gamma function, if we note that by the LaPlace 
transform identity \cite{Andr}
\begin{equation}\label{eq:Ide}
\frac{1}{A^\nu} \,=\, \frac{1}{\Gamma (\nu)}\,\int_0^{\infty}\, {\rm d}s\,{\rm e}^{-s A}\,s^{\nu - 1} \qquad \qquad (A > 0),
\end{equation}
the problem is reduced to the evaluation of a simple gaussian integral. We find, indeed:
\begin{eqnarray}
F(a,0|\nu,2) &\!\!=\!\!& \frac{1}{\Gamma (\nu)}\,\int_0^{\infty}\, {\rm d}s\,{\rm e}^{-s}\,s^{\nu - 1}\,
                                       \int_{-\infty}^{\infty}\, {\rm d}x\, {\rm e}^{-s a x^2}  \\
                      &\!\!=\!\!& \sqrt{\frac{\pi}{a}}\,\frac{\Gamma \left(\nu - \frac{1}{2}\right)}{\Gamma (\nu)}\;. \nonumber
\end{eqnarray}

The same procedures can be exploited to get the analytical expression of the slightly more complicated integral
\begin{eqnarray}\label{eq:Fab}
F(a,b|\nu,2) &\!\!=\!\!& \int_{-\infty}^{\infty}\, {\rm d}x\, \frac{1}{(1 + b x + a x^2)^{\nu}}  \\
                     &\!\!=\!\!& \sqrt{\frac{\pi}{a}}\,\frac{\Gamma \left(\nu - \frac{1}{2}\right)}{\Gamma (\nu)}\,
                                     \left( \frac{1}{1 - \frac{b^2}{4a}}\right)^{\nu -1/2}  \nonumber \\
                      & &               \qquad\qquad  (\Re\,\nu > \frac{1}{2},\,a > 0,\,b^2 < 4a)\;,\nonumber
\end{eqnarray}
and\footnote{We use $\Phi$, instead of $F$, to take into account that the integration is limited to the interval [0, $\infty$) 
and not to the whole real line.}
\begin{eqnarray}
\Phi(a,0|\nu,m) &\!\!=\!\!& \int_0^{\infty}\, {\rm d}x\, \frac{1}{(1 + a x^m)^{\nu}}  \\
                           &\!\!=\!\!& \frac{1}{m\,a^{1/m}}\, \frac{\Gamma \left(\frac{1}{m}\right)\,
                                             \Gamma\left(\nu - \frac{1}{m}\right)}{\Gamma (\nu)} \qquad\qquad (\Re\,\nu > \frac{1}{m},\,a > 0), 
                                             \nonumber
\end{eqnarray}
which can also be recast in terms of the Euler $B$-function \cite{Andr}. It is evident that $m$ can be any real positive number. 

As a simple applications of the previous results we can give an approximate expression for the integral
\begin{equation}\label{eq:Inu}
I_\nu \,=\, \int_{- \infty}^\infty \,{\rm d}x \,\frac{1}{\left[M f(x)\right]^\nu} 
\end{equation} 
where $M > 0$ and $f (x)$ is a real, positive function with a global minimum at $x_0$. In fact, replacing $f(x)$ with its expansion 
around this point up to the second order:
\begin{equation}
f(x) \,\simeq\, f(x_0)\,+\, \frac{1}{2}\,f^{\prime \prime} (x_0)\,(x - x_0)^2 \qquad \qquad (f^{\prime \prime} (x_0) > 0), \nonumber
\end{equation}
the integral \eqref{eq:Inu} assumes the same form of $F(a,b|\nu,2)$ and, thus, according to eq. \eqref{eq:Fab}, we get:
\begin{equation}
I_\nu \,\simeq\, \frac{\sqrt{2 \pi}}{\left[M f(x_0)\right]^\nu}\,\frac{\Gamma \left(\nu - \frac{1}{2}\right)}{\Gamma (\nu)}\,
\sqrt{\frac{f(x_0)}{f^{\prime \prime} (x_0)}}\;. 
\end{equation}
This result is an extension of the steepest descent method \cite{Blei}, usually applied to the evaluation of integrals of the type 
$\int_{- \infty}^\infty \,{\rm d}x \,{\rm e}^{M f(x)}$.
\vspace{0.5cm}

Let us now consider the expansion of the following function:
\begin{equation}\label{eq:Poly}
G(a,b;x|\nu,m) \,=\, \frac{1}{(1 + b x + a x^m)^{\nu}}
\end{equation}
in series of the variable $x$. According to the identity \eqref{eq:Ide}, we find:
\begin{equation}\label{eq:Gint}
G(a,b;x|\nu,m) \,=\, \frac{1}{\Gamma(\nu)}\,\int_0^\infty\, {\rm d}s\, {\rm e}^{-s (1+ b x +  a x^m)}\, s^{\nu - 1}\,.
\end{equation}
By introducing the generalized Hermite polynomials \cite{Datt}
\begin{equation}
H_{n}^{(m)} (x,y) \,=\, n! \, \sum_{r = 0}^{\left[\frac{n}{m}\right]}\, \frac{x^{n - m r}y^r}{(n - m r)!\,r!}\;,
\end{equation}
and their generating function
\begin{equation}
\sum_{n = 0}^{\infty}\, \frac{t^n}{n!}\, H_{n}^{(m)} (x,y) \,=\, {\rm e}^{x t + y t^{m}}\;,
\end{equation} 
the use of the Gamma function properties allows to show that:
\begin{equation}\label{eq:Gabm}
G(a,b;x|\nu,m) \,=\, \sum_{n = 0}^{\infty}\,  \frac{x^n}{n!}\, Q_{n}  (a,b|\nu,m)\;, \qquad(|x| < 1, |b| \le |a|)\;,
\end{equation}
where
\begin{equation}
Q_{n}  (a,b|\nu,m) \,=\, (-1)^{n}\,\frac{n!}{\Gamma(\nu)}\,\sum_{r = 0}^{\left[\frac{n}{m}\right]}\, (-1)^{(m - 1)r}\,
\frac{\Gamma(n - (m - 1) r + \nu)\, b^{n - mr}\, a^{r}}{(n - m r)!\, r!}\;. 
\end{equation}
The polynomials $Q_n$ can also be considered a two-variable generalizations of the ordinary Legendre polynomials \cite{Andr}. 
According to eq. \eqref{eq:Gabm}, a typical incomplete elliptic integral can be written as a series expansion of the $Q$-polynomials
\begin{equation}
\int_0^x \, {\rm d}\xi\, G(a,b;\xi|\nu,m) \,=\, \sum_{n=0}^{\infty}\, \frac{x^{n + 1}}{(n + 1)!}\, Q_{n}  (a,b|\nu,m)\;.
\end{equation}

For the evaluation of integrals over an infinite interval further manipulations are needed, involving the generalization of 
the Euler gamma function defined by the following integral representation
\begin{equation}
\Gamma(x_1,x_2|\nu) \,=\, \int_0^\infty \, {\rm d}t\, {\rm e}^{-x_1 t - x_2 t^2}\, t^{\nu - 1}\;, \qquad \qquad (\Re\,x_2 > 0)\;,
\end{equation} 
whose properties can easily be argued from its defintion. We find:
\begin{equation}
\partial_{x_k} \Gamma(x_1,x_2|\nu) \,=\, - \Gamma(x_1,x_2|\nu - k) \qquad \qquad(k \,=\, 1,2)\;,
\end{equation}
along with the translation property
\begin{equation}\label{eq:G12z}
\Gamma(x_1,x_2 - z|\nu) \,=\, {\rm e}^{z \partial_{x_1}^2}\, \Gamma(x_1,x_2|\nu)\;,
\end{equation}
and the series expansion
\begin{equation}\label{eq:G12}
\Gamma(x_1,x_2|\nu) \,=\, \frac{1}{2}\,\sum_{r = 0}^\infty\, \frac{(-)^r}{r!}\, x_1^r \,x_2^{-\frac{\nu + r}{2}}\, \Gamma\left(\frac{\nu + r}{2}\right)\;.
\end{equation}
As we will discuss later, the above functions can be understood as a two-variable generalization of the Hermite function. 

The extension of the previous generalization of the gamma function to the case
\begin{equation}\label{eq:G1m}
\Gamma(x_1,x_m|\nu;m) \,=\, \int_0^\infty \, {\rm d}t\, {\rm e}^{-x_1 t - x_m t^m}\, t^{\nu - 1}\;, \qquad \qquad (\Re\,x_m > 0)\;,
\end{equation}
is straightforward. In particular, the series expansion in eq. \eqref{eq:G12} is extended in the following way:
\begin{equation}
\Gamma(x_1,x_m|\nu;m) \,=\, \frac{1}{m}\,\sum_{r = 0}^\infty\, \frac{(-)^r}{r!}\, \frac{x_1^r}{x_m^{\frac{\nu + r}{m}}}\, \Gamma\left(\frac{\nu + r}{m}\right)\;.
\end{equation}

According to eqs. \eqref{eq:Poly}, \eqref{eq:Gint}, and the definition \eqref{eq:G1m}, we can write
\begin{eqnarray}
\Phi(a,b|\nu,m) &\!\!=\!\!& \int_0^{\infty}\, {\rm d}x\, \frac{1}{(1 + b x + a x^m)^{\nu}} \,=\,\frac{1}{\Gamma (\nu)}\,\int_0^{\infty}\,
                                            {\rm d}s\, {\rm e}^{-s}\, s^{\nu - 1}\,\Gamma(sb,sa|1;m) \nonumber \\ 
                           &\!\!=\!\!& \frac{1}{m}\, \frac{1}{\Gamma (\nu)}\, \sum_{r = 0}^\infty\, \frac{(-)^r}{r!}\,\frac{b^r}{a^{\frac{1 + r}{m}}}\,
                                            \Gamma\left(\frac{m \nu + (m - 1) r - 1}{m}\right)\,\Gamma\left(\frac{1+r}{m}\right)\;, \nonumber
\end{eqnarray}
which can also be written as
\begin{equation}
\Phi(a,b|\nu,m) \,=\, \frac{\Gamma(a,\hat{b}|\nu;m)}{\Gamma(\nu)}\,,
\end{equation}
where $\hat{b}$ is a kind of umbral notation such that
\begin{equation}
\hat{b}^r \,=\, b^r\, \Gamma\left(\frac{m \nu + (m - 1) r - 1}{m}\right)\,. \nonumber
\end{equation}

The method outlined here can be extended to hyper-elliptic integrals like \cite{Whit}
\begin{equation}\label{eq:Phas}
\Phi({a_s}|\nu;m) \,=\, \int_0^\infty \,{\rm d}x\, \frac{1}{\left(1 + \sum_{s = 0}^m \, a_s x^s\right)^\nu}\;.
\end{equation}
If we limit to the case $m = 3$ the evaluation of the integral requires the introduction of the generalized gamma 
function so defined \cite{Path}:
\begin{equation}
\Gamma(x_1,x_2,x_3|\nu) \,=\, \int_0^\infty\, {\rm d}t\, {\rm e}^{-x_1 t - x_2 t^2 - x_3 t^3}\, t^{\nu - 1}\;,\qquad \qquad (\Re\,x_3 > 0)\;,
\end{equation} 
which can be written in terms of generalized Hermite polynomials as follows
\begin{equation}
\Gamma(x_1,x_2,x_3|\nu) \,=\, \frac{1}{3}\,\sum_{r = 0}^\infty\, \frac{H_r^{(2)}(-x_1,-x_2)}{r! \,x_3^{\frac{r + \nu}{3}}}\, 
\Gamma\left(\frac{r + \nu}{3}\right)\;.
\end{equation}
The above family of generalized gamma functions has extremely interesting properties, that will be argument of a 
forthcoming more detailed investigation. Here we only note the following generalization of eq. \eqref{eq:G12z}
\begin{eqnarray}
\Gamma(x_1,x_2,x_3 + z_3|\nu) &\!\!=\!\!& {\rm e}^{z_3 \partial_{x_1}^3}\, \Gamma(x_1,x_2,x_3|\nu)\;,\nonumber \\
\Gamma(x_1,x_2- z_2,x_3|\nu) &\!\!=\!\!& {\rm e}^{z_2 \partial_{x_1}^2}\, \Gamma(x_1,x_2,x_3|\nu)\;,\nonumber 
\end{eqnarray}
As for the integral in eq. \eqref{eq:Phas}, we find
\begin{equation}
\Phi({a_s}|\nu;m) \,=\, \frac{\tilde{\Gamma}(a_1,a_2,a_3)}{\Gamma(\nu)}\;,
\end{equation}
where
\begin{equation}
\tilde{\Gamma} (a_1,a_2,a_3) \,=\, \frac{1}{3}\,\sum_{n = 0}^\infty \,\frac{\tilde{H}_n^{(2)}(-a_1,-a_2)}{n! \,a_3^{\frac{n + 1}{3}}}\, 
\Gamma\left(\frac{n + 1}{3}\right)\;.
\end{equation}
with
\begin{equation}
\tilde{H}_n^{(2)}(x,y) \,=\, n! \,\sum_{r = 0}^{\left[\frac{n}{2}\right]}\, \frac{x^{n - 2 r}y^r}{(n - 2 r)!\,r!}\,
\Gamma\left(\frac{2 n + 3 (\nu - r) - 1}{3}\right)\;.
\end{equation}
\vspace{0.5cm}

Let us now introduce the Hermite functions, defined in the following way \cite{Lebe}:
\begin{equation}\label{eq:Herm}
H_\nu (x) \,=\, \frac{1}{\Gamma( - \nu)}\, \int_0^\infty\, {\rm d}t\, {\rm e}^{-2 x t - t^2}\, t^{\nu - 1}\;, \qquad\qquad (\Re \nu > 0)\;.
\end{equation}
The functions we have introduced before are, for $m > 2$, a generalization of the standard form given in eq. \eqref{eq:Herm}. In the 
case $m = 2$ we find
\begin{equation}\label{eq:29}
\Gamma(x_1,x_2|- \nu) \,=\, \Gamma(-\nu)\,x_2^{-\frac{\nu + 2}{2}}\,H_\nu \left(\frac{x_1}{2 \sqrt{x_2}}\right)\,.
\end{equation}

As for the Hermite functions with $m > 2$, it is worth considering the case when $\nu$ is an integer. We set, therefore
\begin{eqnarray}\label{eq:30}
H_\nu^{(m)} (x_1, x_m) &\!\!=\!\!& \frac{1}{\Gamma(-\nu)}\,\Gamma(x_1,x_m|-\nu;m) \\
                                          &\!\!=\!\!&\frac{1}{\Gamma(-\nu)}\,\int_0^\infty \, {\rm d}t\, {\rm e}^{-x_1 t - x_m t^m}\, t^{- \nu - 1}\;,
                                          \nonumber 
\end{eqnarray}
and note that
\begin{equation}\label{eq:31}
H_{-1}^{(m)} (x_1, x_m) \,=\, \int_0^\infty \, {\rm d}t\, {\rm e}^{-x_1 t - x_m t^m}\;.
\end{equation}
which can be exploited to generate all the other orders according to the identity
\begin{equation}\label{eq:32}
H_{-(n +1)}^{(m)} (x_1,x_m) \,=\, \frac{(-)^n}{n!}\, \partial_{x_1}^n \,H_{-1}^{(m)} (x_1,x_m)\;.
\end{equation}
It is worth to stress also the following generating function
\begin{equation}\label{eq:33}
\sum_{n = 0}^\infty \, z^n\, H_{- n - 1}^{(m)} (x_1,x_m) \,=\, H_{-1}^{(m)} (x_1 - z, x_m) \;.
\end{equation}
As it is well known, in the standard case ($m = 2$) $H_{-1}^{(m)} (x_1, x_m)$ can be written in terms of the error function.
\vspace{0.5cm}

In this letter we have given an idea of the wealth of implications associated with the techniques we have briefly illustrated. 
In a forthcoming paper we will draw further and general conclusions.


\begin{thebibliography}{9}

\bibitem{Andr}
L. C. Andrews, 
\emph{Special functions for Applied Mathematicians and Engineers},
McMillan, New York, 1985.

\bibitem{Blei}
N. Bleistein and R. Handelsman, 
\emph{Asymptotic Expansion of Integrals},
Dover, New York, 1975.

\bibitem{Datt}
G. Dattoli, A. Torre, and M. Carpanese, 
J. Math. Anal. Appl. \textbf{227} (1998), 98. 

\bibitem{Whit}
E. T. Whittaker and G. N. Watson,
\emph{A Course in Modern Analysis}, 
Cambridge University Press, Cambridge, 1990.

\bibitem{Path}
M. A. Pathan, 
Scientia A: Mathematical Science \textbf{12} (2006), 9. 

\bibitem{Lebe}
N. N. Lebedev 
\emph{Special Functions and their Applications},
Dover, New York, 1972.

\end{thebibliography}
\end{document}